\documentclass[twocolumn,showpacs,preprintnumbers,amsmah,amssymb]{revtex4}
\usepackage{graphicx}
\usepackage{dcolumn}
\usepackage{bm}

\newcommand{\angstrom}{A \kern -.4em\raise 1.4ex \hbox{\tiny °}}
\newcommand{\strikline}[1]{#1 \kern -0.5em\raise 0.5ex \hbox{--}}

\begin{document}                


\title[]{Quantitative description of the azimuthal dependence of the exchange bias effect}

\author{Florin Radu\footnote{florin.radu@bessy.de}\footnote{Present address: BESSY GmbH, Albert-Einstein-Str.
15, D-12489 Berlin, Germany.}, Andreas Westphalen, Katharina
Theis-Br\"ohl, and Hartmut Zabel}
\address{Institut f\"{u}r
Experimentalphysik/Festk\"{o}rperphysik, Ruhr-Universit\"{a}t
Bochum, D-44780 Bochum, Germany}
\date{\today}

\begin{abstract}

While the principal features of the exchange bias between a
ferromagnet and an antiferromagnet are believed to be understood, a
quantitative description is still lacking. We show that interface
spin disorder is the main reason for the discrepancy of model
calculations versus experimental results. Taking into account spin
disorder at the interface between the ferromagnet and the
antiferromagnet by modifying the well known Meiklejohn and Bean
model, an almost perfect agreement can be reached. As an example
this is demonstrated for the CoFe/IrMn exchange biased bilayer by
analyzing the azimuthal dependence of magnetic hysteresis loops from
MOKE measurements. Both, exchange bias and coercive fields for the
complete 360$^\circ$ angular range are reproduced by our model.

\end{abstract}
 \maketitle

The exchange bias system refers to the shift of the ferromagnetic
(F) hysteresis loop to positive or negative values when the F
system is in contact with an antiferromagnetic (AF) system and
cooled in an applied magnetic field through the N\'eel temperature
of the AF system. The exchange bias (EB) phenomenon is associated
with the interfacial exchange coupling between ferromagnetic and
antiferromagnetic spin structures,  resulting in a unidirectional
magnetic anisotropy~\cite{bean:1956}. While the unidirectional
anisotropy was successfully introduced by Meiklejohn and Bean
(M\&B), the origin of the enhanced coercive field is yet not well
understood. The details of the EB effect depend crucially on the
AF/F combination chosen and on the structure and thickness of the
films~\cite{berkowitz:1999,nogues:1999}. However, some
characteristic features apply to most systems: (1) $H_{EB}$ and
$H_c$ increase as the system is cooled in an applied magnetic
field below the blocking temperature $T_B \leq T_N$ of the AF
layer, where $T_N$ is the N\'{e}el temperature of the AF layer;
(2) the magnetization reversal can be different for the ascending
and descending part of the hysteresis
loop~\cite{fitzsimmons:2000,radu:2002:1,gierlings:2002,lee:2002,radu:2003:1};
(3) thermal relaxation effects of $H_{EB}$ and $H_c$ indicate that
a stable magnetic state is reached only at very low
temperatures~\cite{geoghegan:1998,goodman:2000, radu:2002:2}.

Several theoretical models have been developed for describing
possible mechanisms of the EB effect, including domain formation
in the AF layer with domain walls perpendicular to the AF/F
interface~\cite{malozemoff:1987}, creation of uncompensated excess
AF spins at the interface~\cite{schulthess:1998}, or the formation
of domain walls in the AF layer parallel to the
interface~\cite{mauri:1987,kim:2005}. Another approach is the
consideration of diluted antiferromagnets in an exchange field. In
the work described in Ref.~\cite{milt:2000,keller:2002,nowak:2002}
the discussion about compensated versus uncompensated interfacial
spins is replaced by a discussion of net magnetic moments
\textit{within} the antiferromagnetic layer. Depending on the
complexity of the system, the models can explain some but not all
features of experimental hysteresis loops. Here we provide a new
model which can describe all features, including the azimuthal
dependence of $H_{EB}$, $H_{c}$, and the AF thickness dependence.
In this letter we concentrate on the azimuthal dependence (AD),
the thickness dependence will be reported
elsewhere.

The AD of $H_{EB}$ and $H_{c}$ is an important feature of all EB
systems. First experiments were performed on NiFe/CoO
bilayers~\cite{ambrose:1997}, where it was suggested that it can
be best described by a cosine series expansions with odd and even
terms for $H_{EB}$ and $H_c$, respectively. A recent study of
IrMn/CoFe bilayers\cite{outon:2005} showed that even so good
agreement between the data and the simulations based on the cosine
functions can be achieved, still some disagreement exists. In an
another approach, Mewes et al.~\cite{mewes:2002} showed that the
AD of $H_c$ and $H_{EB}$ can be well described within the
Stoner-Wohlfarth model. However the magnitude of the coercivity is
not explained.

We have measured the AD of  CoFe/IrMn exchange bias bilayers via
longitudinal (L) and transverse (T) magnetization curves. The
experimental data are described by a modified M\&B model assuming
the existence of a spin disorder (SD) at the F/AF interface. In
the F/SD/AF system the SD layer has the role to reduce the
EB-field and to mediate coercivity from the AF to the F layer. The
formation and existence of the SD layer has been shown in
Ref.~\cite{radu:2003:1,Radu:xrms,roy:2005,gruyters:2005}. We
believe that it is an essential feature of all EB systems.

Exchange-biased F/AF
polycrystalline~\cite{schmalhorst:2003,hoink:2005} bilayers   $
Ir_{17}Mn_{83} (15 \, nm)/Co_{70}Fe_{30}(30 \, nm)$ were prepared
by magnetron sputtering on $Si/SiO2/Cu(30 \, nm)$ substrates,
covered by a $ Ta(5 \, nm)$  protection layer. The base pressure
was below $1\,\times \, 10^{-7}$ Torr at an Ar pressure of $3 \,
\times \, 10^{-3}$ Torr. The initial EB direction is set by an
annealing step after deposition for 1 h at 548~K which is higher
then the blocking temperature\cite{nogues:1999} of exchange bias
systems having IrMn layer as the AF layer. The annealing  magnetic
field $H_{ann}=1$~kOe was applied and maintained parallel to the
film plane ~\cite{hoink:2005}.

\begin{figure}[ht]
\begin{center}
\includegraphics[clip=true,keepaspectratio=true,width=1\linewidth]{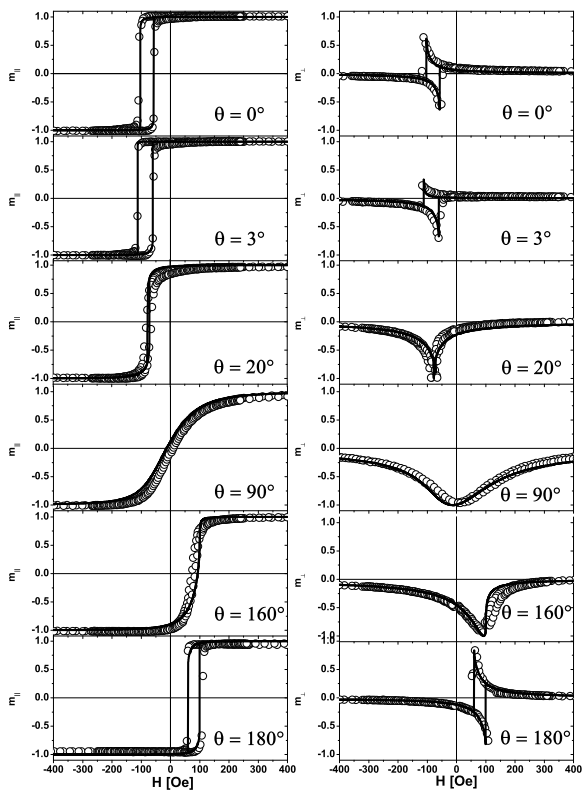}
\caption{\label{fig1} Experimental (open circles)and simulated
hysteresis loops (black lines) for different azimuthal angles. The
simulated curves are calculated by the Eq.~\ref{SG2} with the
following parameters: $f=80\% , \, R=5.9/f, \, \gamma= 20^\circ$.}
\end{center}
\end{figure}

\begin{figure}[!ht]
\begin{center}
\includegraphics[clip=true,keepaspectratio=true,width=1\linewidth]{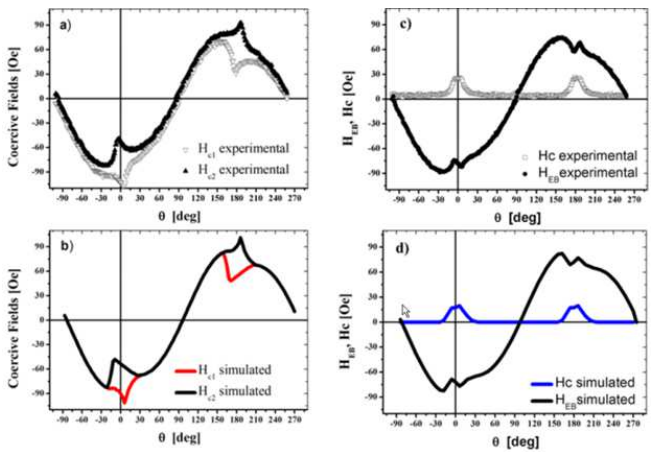}
\caption{\label{fig2} a) Azimuthal dependence of the  coercive
fields $H_{c1}$ (filled symbol) and $H_{c2}$ (open symbols)
extracted from the experimental hysteresis loops. At $\theta=0,
180^\circ$, which corresponds to the  field cooling direction the
coercive fields deviate one from each other within a 20$^\circ$
angular range. b) Calculated coercive fields $H_{c1}$ and $H_{c2}$
 as a function of the azimuthal angle, using Eq.~\ref{SG2} with the following
parameters: $f=80\% , \, R=5.9/f, \, \gamma= 20^\circ$.  c) The
experimental coercive field and exchange bias field as a function
of the azimuthal angle $\theta$. The field cooling orientation
corresponds to $\theta=0$. bd) Simulated coercive field and
exchange bias field  as a function of the azimuthal angle. The
curves are delivered by the Eq.~\ref{SG2} with the following
parameters: $f=80\% , \, R=5.9/f, \, \gamma= 20^\circ$ }
\end{center}
\end{figure}

The sample was measured using a vector-MOKE
setup~\cite{westphalen:2005}.  A number of 360 pairs of L- and T-
components of the magnetization ($m_{||}$ and $m_{\perp}$
respectively) were measured for an external field orientation with
respect to the field cooling direction ranging from $0^\circ$ to
$360^\circ$. All loops were taken  at room temperature. In this
geometry, the sample is kept fixed during the measurements,
whereas the orientation of the applied external field was varied.
Characteristic L- and T-magnetization curves are shown in
Fig.~\ref{fig1}. In order to observe fine variations of the EB
field, the one degree increment of the azimuthal angle is
required. The L-hysteresis loops were used to extract the coercive
fields $H_{c1}$ and $H_{c2}$ as shown in Fig.~\ref{fig2}a, which
further provides the coercive field $H_c=(-H_{c1}+H_{c2})/2$ and
the EB field $H_{EB}=(H_{c1}+H_{c2})/2$ as plotted in
Fig.~\ref{fig2}c.

We first discuss  the experimental observations in
Figs.~\ref{fig1},~\ref{fig2}a,~and~\ref{fig2}c. In Fig.~\ref{fig1}
some representative L- and T-loops are shown. A distinct feature
of the system is  the magnetization reversal which occurs via
coherent rotation~\cite{paul:2004} as seen from the non-vanishing
transverse loops, which are remarkably well reproduced by
numerical simulations(solid lines), to be discussed further below.
The maximum EB field $H_{EB}=90~Oe$ is achieved at $\theta \approx
-20^\circ$ (see Fig.~\ref{fig2}c and \ref{fig2}d), where $\theta$
is the azimuthal angle with respect to the field cooling direction
defined as $\theta=0$. This off-angle is one of the salient
features reported here, as usually the maximum of the EB field is
believed to occur parallel to the field cooling direction.

The L-loop at $\theta=0^\circ$ is completely symmetric. This is
also seen in the transverse magnetization, where the forward and
reverse components have the same magnitude, but are doubly
mirrored with respect to $m_\perp=0$ and $H_{EB}$.
This is not always the case. Only a few degrees forward at
$\theta=3^\circ$ both,  L and T-loops become asymmetric. The
forward branch of the L-loop is steep, while the reverse branch is
more rounded. It is remarkable that within an azimuthal angle of
only $3^\circ$ such a strong asymmetry of the L-loop develops.
This asymmetry is different from the one observed due to the
training effect~\cite{radu:2002:1,radu:2003:1}. The former is
completely reversible while the latter is not.

As the azimuthal angle is further increased, $H_{c}$ disappears at
about $\theta=20^\circ$ and reappears again in a symmetrical
fashion close to $\theta \approx 160^\circ$. The vanishing $H_c$
can be understood from the T-loops, where it is clearly seen that
the rotation of the ferromagnetic spins do not make a complete
$360^\circ$ rotation, but almost reversibly rotate within the half
circle of $180^\circ$. Therefore the angle of the magnetization
orientation, from which the coercive fields are extracted, takes
the same value for both $H_{c1}$ and $H_{c2}$.

In Fig.~\ref{fig2}a the coercive fields are plotted versus the
azimuthal angle $\theta$. We notice that, globally, they follow
the expected unidirectional behavior~\cite{bean:1956}, but some
striking deviations are recognizable. In particular,  close to the
field cooling direction spike like features appear. While the
coercive fields $H_{c1,c2}$ are well behaved for most of the
angles, this is not the case for field directions close to the
field cooling orientation. Here $H_{c1}$ and $H_{c2}$ deviate one
from each other, resulting in a non-vanishing coercive field as
seen in Fig.~\ref{fig2}c. Its maximum value ($H_c^{MAX}=20$~Oe) is
about four times lower than the maximum value of the exchange bias
field. Finite values are observed within a $20^\circ$ degree range
centered at $\theta=0^\circ$ and $\theta=180^\circ$ and almost
vanish outside this range.

The $H_{EB}$ dependence on the azimuthal angle (solid symbols in
Fig.~\ref{fig2}c) clearly shows the unidirectional anisotropy. In
addition a peculiar and sharp modulation is seen with a low
amplitude. These features appear close to the field cooling
orientation and also close to the opposite orientation. They
cannot be reproduced satisfactorily with the empirical description
based on a cosine series expansion as suggested in
Ref.~\cite{ambrose:1997}. Therefore we need a more realistic
model, which is discussed next.

The M\&B model~\cite{bean:1956} assumes that the AF spins rigidly
form an AF state, but they may slightly rotate as a whole during
the magnetization reversal of the F layer. Within the M\&B model,
enhanced coercivity is not accounted for. The interface is assumed
to be perfectly uncompensated with the interface AF spins having
the same anisotropy as the bulk spins. However, the interface is
never perfect. Roughness, deviations from stoichiometry,
interdiffusion, structural defects, low spin coordination at
surface sites~\cite{kodama:1997}, etc. cause non-ideal magnetic
interfaces. It is therefore natural to assume that, on average, a
fraction of the AF spins have lower anisotropy as compared to the
bulk ones.  These interfacial AF spins can rotate together with
the ferromagnet~\cite{ohldag:2003,Radu:xrms,roy:2005}. They
mediate the exchange coupling, induce an enhanced coercivity, but
soften the extreme coupling condition assumed by M\&B. Therefore,
we assume that the anisotropy of the AF interface layer varies
from $K_{int}=0$ next to the F layer to $K_{int}=K_{AF}$ next to
the AF layer, where $K_{AF}$ is the anisotropy constant of a
presumably uniaxial antiferromagnet. This anisotropy gradient
across the interface governs the enhanced anisotropy of the
ferromagnetic layer, which otherwise would be close to zero for
CoFe. So far it was believed that the enhanced coercivity in F/AF
exchange biased systems is caused by compensated AF spins at the
F/AF interface. We argue that for most of the AF materials a
compensated or uncompensated spin having the same anisotropy as
the bulk AF layer would be practically impossible to reverse by
rotating the F layer. Therefore we need to assume low anisotropy
AF spins in order to quantitatively describe the experimental
data.

A direct indication of the rotating AF spins is revealed by soft
X-ray magnetic dichroism~\cite{ohldag:2003,Radu:xrms,roy:2005}.
Element specific hysteresis loops show that some spins belonging
to the AF layer rotate reversibly with the F spins. Due to the
shift of the hysteresis loop it is obvious that another part of
the AF layer is frozen. Therefore the AF layer can be considered,
to a first approximation, as consisting of two types of AF states.
One part having a large anisotropy preserving the AF state, and
another interfacial part with a weaker anisotropy, allowing the
spins to rotate together with the F spins. Moreover, polarized
neutron scattering~\cite{radu:2002:1,radu:2003:1} revealed two
further effects related to the magnetic state of the CoO/Co
interface (which is similar to the CoFe/IrMn
system~\cite{mccord:2003}) during the magnetization reversal: a)
the interface is disordered containing domains and domain walls
even in saturation, similar to a spin-glass  system; b) the
interfacial ferromagnetic spins are not collinear with the applied
field direction during the reversal.

The experimental results mentioned above are in our model
accounted for by two averaging interface properties: (a) the
existence of low AF anisotropy spins,  to which we assign an
effective average anisotropy $K^{eff}_{SD}$; (b) a
non-collinearity angle $\gamma$. Adding these two parameters, the
modified M\&B model reads:
\begin{eqnarray}
E=&-&\mu_0\, H \, M_F \, t_F \, \cos(\theta-\beta)+K_F \,t_F \,
\sin^2(\beta) \nonumber\\
 &-&\mu_0\, H \, M_{SD} \, t_{SD} \, \cos(\theta-\beta)+K^{eff}_{SD} \, \sin^2(\beta-\gamma) \nonumber\\
 &+& K_{AF} \, t_{AF}\, \sin^2(\alpha)\nonumber\\
 &-&J_{EB}^{eff} \,\cos(\beta-\alpha), \label{SG1}
\end{eqnarray}
where $J_{EB}^{eff}$ is the reduced interfacial exchange energy,
the $\gamma$ ($\gamma\ge0$) angle is the averaged angle of the
effective SD anisotropy which can be considered as
 (partially) fanning  in orientation with respect to the average anisotropy
orientation of the AF layer~\cite{macedo:2004},
 $\alpha$ is the
average angle of the AF uniaxial anisotropy~\cite{bean:1956},
$M_{AF}$ is the magnetization of the SD interface, and  $t_{SD}$
is the SD interface thickness. For
 simplifying the numerical analysis we neglect the  $-\mu_0 \, H\, M_{SD}\, t_{SD}$
term, because $M_{SD}\, t_{SD}$ is  small. Furthermore we neglect
the crystal anisotropy of the ferromagnetic layer ($K_F=0$), which
is  well justified because Co$_{70}$Fe$_{30}$ is a soft magnetic
material with a coercivity in the range of few
Oersteds~\cite{muhge:1995}.


The interface anisotropy, which leads to enhanced coercivity,
characterizes the quality of the interface. When $K^{eff}_{SD}$ is
zero, the system behaves ideally as described by the M\&B
model~\cite{bean:1956}, i.e. the coercive field is zero and the
exchange bias field is finite. In the other case when the
interface is disordered, we relate the  effective SD anisotropy to
the available interfacial coupling energy as follows:
\begin{eqnarray}K^{eff}&=& (1-f)\, J_{EB}\nonumber \\
J_{EB}^{eff}&=&f \, J_{EB}, \label{none}
\end{eqnarray}
where $J_{EB}$ is the total exchange energy of an ideal system
without additional coercivity. With this assumption the absolute
value of the EB field is reduced by the factor $f$ as compared to
the M\&B model. The factor $f$ describes the conversion of
interfacial energy into coercivity through  rotation of
interfacial AF spins.

Next, we write the system of equations resulting from the
minimization of the Eq.~\ref{SG1} with respect to the $\alpha$ and
$\beta$ angles:
\begin{eqnarray}
&h& \sin(\theta-\beta) +\frac{(1-f)}{f} \, \sin(2\,
(\beta-\gamma))+  \sin(\beta-\alpha)=0 \nonumber  \\
&R& \sin(2\, \alpha)- \sin(\beta-\alpha)=0, \label{SG2}
\end{eqnarray}
where, $h=H / \, [-J_{EB}^{eff}/(\mu_0 \, M_F \, t_F \,)] \,$ is
the reduced field,  and $R = K_{AF}\,t_{AF}/J_{EB}^{eff}$ is the
R-ratio defining the strength of the AF layer. This system of
equations can easily be  solved numerically, but it cannot deliver
a simple analytical expression for the exchange bias. Numerical
evaluation provides the angles $\alpha$ and $\beta$ as a function
of the applied magnetic field $H$. The reduced L- and T-
components of the magnetization are $m_{||}=\cos(\beta)$ and
 $m_{\perp}=\sin(\beta)$, respectively. These are the two observables measured by
vector-MOKE. Note that the anisotropic magnetoresistance (AMR) and
PNR hide the chirality of the ferromagnetic spin rotation as they
provide $\sin^2(\beta)$ information, whereas MOKE reveals the
chirality through $\sin(\beta)$ information.

In Fig.~\ref{fig1} calculated magnetization components are plotted
together with the experimental data points, and in
Figs.~\ref{fig2}b~and~\ref{fig2}d the azimuthal dependence of the
coercive fields and exchange bias field are plotted and compared
to the experimental data in Fig.~\ref{fig2}a~and~\ref{fig2}c. In
all cases we find an astounding agreement between calculated
curves and experimental data. It is remarkable, that the AD of the
EB field and the coercive fields ($H_{c1}, H_{c2}, H_c)$ are
completely reproduced by the SD model. The parameters used are:
$f=(80~\pm 2)\, \%$, $\gamma=(20~\pm 2)^\circ$ and $R=5.9/f$. For
calculating the value of  the R-ratio we used the anisotropy
constant ($K_{AF}$) measured in Ref.~\cite{steenbeck:2004}. The
conversion factor is related to the magnitude of the coercive
field with respect to the shift of the loop. The $\gamma$ angle
plays an important role. It represents the mean angle of the spin
disorder at the interface with respect to $\theta=0$. For
instance, when $\gamma$ is zero, the coercive field is much
enhanced at $\theta=0, 180^\circ$ as compared to the experimental
data, and the azimuthal dependence of $H_{EB}$ and $H_c$ cannot be
reproduced.

Our SD model also describes the AF thickness dependence of the EB.
Notably the peak like behavior close to the critical AF thickness
is reproduced by Eq.~\ref{SG2} \cite{radu:phd} , as has been
recently observed for IrMn/Co heterostructures~\cite{ali:2003}.
The key parameters for achieving the enhanced exchange biased are
the conversion factor $f$ and the $R$-ratio. When the $R$-ratio is
close to critical value of one, the AF spins rotates to higher
$\alpha$ angles ($\alpha < 45^\circ$) during the magnetization
reversal. Due to the intrinsic asymmetric reversal, the AF layer
absorbs also asymmetrically some fraction of the coercive fields
$H_{c1}$ and $H_{c2}$ and give rise to an increased EB field. A
reduced $R$-ratio can be achieved by either reducing the thickness
of the AF layer or by reducing its anisotropy.  A condition for
the peak in exchange bias to occur close to the critical values of
the $R$-ratio is  the conversion factor to be smaller then approx.
f=0.85.  A reduced  $f$ factor can be achieved also at elevated
temperatures through the thermal fluctuations affecting disordered
interface.

In conclusion, taking into account spin disorder at the interface
between the AF and F layer we have achieved a compelling agreement
between the azimuthal dependence of the coercivity and exchange
bias field in the IrMn/FeCo systems. The new key physical concept
is a realistic state of the interface characterized by a reduced
AF anisotropy. This disorder governs the enhanced coercivity in
the ferromagnetic layer and reduces the exchange bias field to
realistic values. We believe that this is a general feature of EB
systems. By controlling the degree of spin disorder and the
thickness of the interfacial layer, better control over the
exchange bias of magnetic heterostructures could be achieved.

We would like to thank  J. Schmalhorst, V. H\"oink, and H.
Br\"uckl from the University of Bielefeld for providing the
samples. We gratefully acknowledge support through the
Sonderforschungsbereiche 491 "Magnetische Heteroschichten:
Struktur und elektronischer Transport" of the Deutsche
Forschungsgemeinschaft.

\end{document}